# Impact of Network and Host Characteristics on the Keystroke Pattern in Remote Desktop Sessions

Ievgeniia Kuzminykh[1], Bogdan Ghita[2], and Alexandr Silonosov[3]

*Abstract*—Authentication based on keystroke dynamics is a convenient biometric approach, easy in use, transparent, and cheap as it does not require a dedicated sensor. Keystroke authentication, as part of multi factor authentication, can be used in remote display access to guarantee the security of use of remote connectivity systems during the access control phase or throughout the session. This paper investigates how network conditions and additional host interaction may impact the behavioural pattern of keystrokes when used in a remote desktop application scenario. We focus on the timing of adjacent keys and investigate this impact by calculating the variations of the Euclidean distance between a reference profile and resulting profiles following such impairments.

The experimental results indicate that variations of congestion latency, whether produced by adjacent traffic sources or by additional remote desktop interactions, have a substantive impact on the Euclidian distance, which in turn may affect the effectiveness of the biometric authentication algorithm. Results also indicate that data flows within remote desktop protocol are not prioritized and therefore additional traffic will have a significant impact on the keystroke timings, which renders continuous authentication less effective for remote access and more appropriate for one-time login.

*Index Terms*—biometric pattern, continuous authentication, Euclidean distance, keystroke dynamics, MFA, remote display protocol, RDP.

## I. INTRODUCTION

THE ubiquity and speed of Internet access led over the past decade to an exponential increase in the use of cloud computing and virtualized services, including Desktop-as-a-Service (DaaS), both taking advantage of the ability to remotely provide computing resources, from single systems to entire infrastructures. In this context, Virtual Desktop Infrastructure (VDI) and remote display access allows users to reach their allocated hosts from a terminal with reduced processing, memory, or/and storage capabilities. Such a terminal, called a thin client, renders the graphical user interface of the remote applications and transmits mouse and keyboard events from thin client to the server. The network communication protocols used to transfer the screen image, audio, mouse events and typed keystroke data are referred to as remote display protocols. The use of these protocols is not limited to accessing the remote host, but also for remote administration of servers and workstations, for providing access during conferences, demonstrations, or collaborative work. Remote administration is one of the core activities for IT administrators today and Remote Desktop is the administrator's go-to tool. Remote Desktop connections are essential for efficient remote management and troubleshooting. Remote display protocols are a common target for man-in-the-middle cyberattacks, due to their widespread use and the fact that such connections often have elevated privileges because they are typically used to perform administrative tasks. This context makes remote desktop security risks a top concern for network administrators, security experts, and analysts.

The typical option to provide security is using a network access control method, provided through the identification and authentication of the users. In order to enhance its effectiveness, user authentication may use multi-factor techniques or the inclusion of additional user identification steps. Behavioral biometrics, based on free-text keystroke dynamics, tend to be the de-facto alternative, as they are promising enough to achieve cost effective continuous authentication system.

Due to their unique characteristics, typing keystroke dynamics can be used to create user identification templates for authentication. The keystroke dynamics algorithm computes a user biometric pattern by using a set of temporal metrics derived from keystroke events generated by computer user throughout a session [1]. Continuous user authentication based on keystroke dynamics recognition has raised a lot of interest among researchers [2,3]. Many studies report a high recognition rate, and this forms a condition to implement it for remote display connection.

A continuous authentication system based on keystroke dynamics and used in VDI may be considered a latency-sensitive application since the computer display content, the pressed keystrokes and the mouse movements are being transferred by remote desktop protocols in a similar time-sensitive fashion as a video streaming service. During the transmission of data from the thin client over remote display protocol, the temporal metrics of user input might be affected by the characteristics and quality of communication network. Therefore, it is relevant and necessary to experiment whether degraded network conditions may affect the accuracy of the continuous authentication system when keystroke dynamics are sent over to the remote VDI display.

The relationship between latency-sensitive applications and the Mobile Edge Computing architecture was summarized in

[1] King's College London, UK (e-mail: ievgeniia.kuzminykh@kcl.ac.uk)
[2] University of Plymouth, UK
[3] Blekinge Institute of Technology, Karlskrona, Sweden

[4], with the authors proposing measures such as network monitoring, efficient routing mechanisms, and Cloudlets to address the existing issues. In addition, at the transport layer, Ferlin et al. [5] proposed an improvement to the TCP and MTCP loss recovery mechanisms in high delay and lossy networks in order to minimize round-trip times loss recovery for latency-sensitive applications, such as video-streaming. In addition, the impact of network latency on remote desktop sessions was studied in [6,7]. As part of the study, the authors demonstrated that user experience and human visual perception are affected by the VDI technology during capture-transfer-render phases. Virtual environment represents a specific context in cybersecurity dimension of behavioral biometric.

In our study, we measure the impact of the network characteristics and background running applications onto the resulting behaviour pattern based on keystroke dynamics that are transmitted in the Virtual Desktop Infrastructure by using a thin client with a Microsoft RDP cloud access agent. The experimental tests replicate the effect of network conditions on the packets transmitted between a thin client and a remote desktop server, using a number of network- and system-based applications to emulate concurrent network activity. This environment replicates a typical VDI interaction, including emulating the effect of cross-traffic and congestion. We acknowledge that RDP is not the dominant market player but it provides a simplified, free setup and, more importantly, it shares functionality principles with other VDI protocols.

Given the variety of existing keystroke authentication solutions, we aim to observe the generic impact of network variations onto the resulting keystroke pattern rather than evaluate individual biometric authentication solutions. To follow, a secondary aim of the study is to inform developers of biometric solutions to consider the network impact on the keystroke pattern before getting into machine learning because this modified by the network keystroke pattern can cause make the wrong decision by machine. In this context, keystroke-based user identification and authentication are outside the scope of this paper but considered for future work or integration with enhancement of biometric techniques.

## II. ARCHITECTURE OF REMOTE DESKTOP APPLICATION

From an architectural perspective, a remote desktop application includes a thin client and a remote display server (see Fig.1). The thin client is typically a workstation with a remote desktop application, and the server is the host where the user application(s) reside/run. The remote user logs in via different remote display protocols, including Microsoft RDP, Teradici PCoIP, or Citrix ICA. For the purpose of this paper, a keystroke acquisition component was added on the server for data collection. The authentication scheme can be implemented on top of the application using its two components, with data collection and processing on the server side.

During the enrolment phase, the collected data allows creating a reference profile that is stored in a database. Subsequently, every time a user either logs in or interacts with a remote session, their keystrokes are collected on the server side and compared with the reference profile. In the case of a multi-user environment (Windows Server2012, 2016), if multiple users are logged in, each user reference profile is associated within their corresponding individual session.

Remote display protocols capture input events from locally attached keyboard, transfer them to remote desktop over the network via TCP/IP and replay or render input events on a remote display to simulate user input. The additional cross-traffic and congestion affect the timing of the received keyboard events. However, the keystroke events sent by the VDI software, for example Microsoft RDP, does not contain timestamps[1] [8]. TCP does indeed eliminate misordering through the in-sequence delivery, but time spacing information is not preserved through the transmission. While protocols such as PCoIP for VMWare [9] or ICA for Citrix [10], may differ in specification from RDP, it is important to note that they all share the lack of timing information for keystroke events.

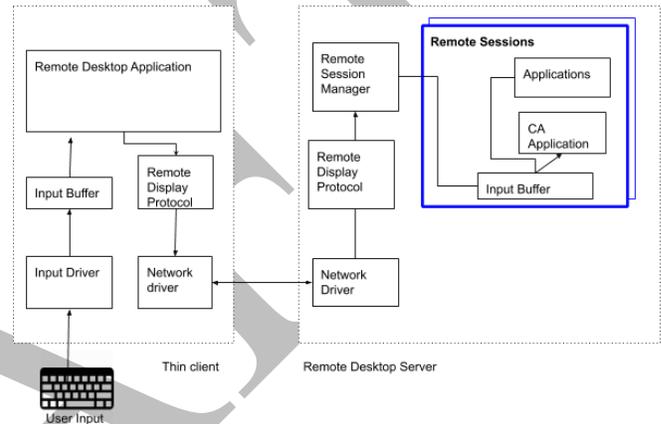

Fig.1. Virtual desktop infrastructure.

Another characteristic of the VDI environment is that authentication and user access control are performed on the remote system, which means that keystroke patterns are captured and verified on the remote session computer rather than on the thin client computer, as shown in Fig.1. This is clearly not optimal from a security perspective, but it is in place for two reasons. Firstly, the thin client has reduced/minimal functionality, therefore adding processing tasks would be against its architecture design. Secondly, the assumption made by VDI software implementations is that the initial authentication is sufficient for ensuring the security of the application. In addition to the processing required for the applications they run, remote display protocols also may use significant CPU resources to detect screen changes and compress screen image and audio data.

---

[1]https://docs.microsoft.com/en-us/openspecs/windows_protocols/ms-rdpbcgr/1e4ac25c-a035-4c91-8341-88424945a4c8

CPU usage varies depending on entropy of screen image changes [11]. In addition, the remote desktop server software in a cloud environment may share CPU resources among multiple remote sessions. Thus, on a low performance or dynamically allocated VDI system, the behaviour and timing of some of the concurrent remote sessions may affect the performance of the other sessions, which in turn will affect user-mode applications.

Given VDI requirements, keystroke-based identification and authentication are well suited for transparent authentication in remote desktop applications, as it does not require specialized equipment and the login credentials of each user can be also used for biometric authentication. While its inherent noise characteristics make it a less suitable candidate for single factor authentication, it represents the ideal option for continuous authentication throughout the session, especially in the case of a session hijack attack.

Biometric identification systems typically follow three processing steps: acquisition, template generation and matching.

1. Acquisition: raw data is captured with a respective technique.

2. Template generation: the acquired scanned content must be stored for procession. On this stage the filtration is performed on extraneous information stored content to isolate the uniquely differentiating characteristics of the biometric attribute, and the template is created. The template is a mathematical representation of a biometric data and often refers to user profile.

3. Matching: the processed sample is cross-compared with the stored reference template in the database. Identification and authentication are successful if a match is found, else the user is classified as an impostor.

During the acquisition step, the raw data is recorded in real time, while the respective user types, as a collection of timing parameters, which are then processed and converted into attributes, to be used for authenticating or identifying the user. From the intrusiveness perspective, there are two types of user input when performing keystroke dynamics: fixed- and free-text. In fixed-text authentication, a user needs to type a predefined phrase in order to have their biometric characteristics collected and evaluated, and subsequently be allowed to access system; the most common form of this type of data is when an user enters their login ID and password. In free-text authentication, the user may type any text, including command keys, hotkeys, arrows, which makes the method ideal for continuous authentication to be run transparently in the background while the user is interacting with the system. In this case, the user enters arbitrary text as desired or a long text string of 500–1000 characters. The reference model, or template associated with the user, is created on the second step and reflects the unique style of typing. It is stored for subsequent authentication, and would require entering text either several times if a fixed-text input form is used or once when using the free-text form.

The most significant attributes extracted when a person types are digraph-specific (see Fig.2): the time interval between pressing and releasing a specific key (dwell time), and the time between releasing and pressing two successive keys "AB" (flight time). These functions, along with several others, are used to build a model of how a person types. While it is not designed to be the main authentication technique, the use of a biometric method provides an additional level of security for the system since it would not be sufficient for the attacker to have the access password, they would also be required to enter it in the same way the valid user entered it during enrolment and user profile acquisition step. If not, the login attempt is rejected.

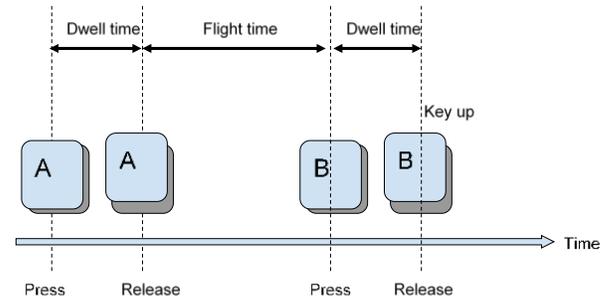

Fig.2. Keystroke dynamics attributes.

Other attributes can also be captured [12], they could be device-based or network-based. Examples of features for keystroke dynamics (during entering to device/system) extraction are file system and registry, mouse dynamics, the duration of password typing, the use of "invisible keys", system process, browser, flow-based features.

### III. LITERATURE REVIEW

Prior research in remote display and thin client functionality highlighted that network latency and virtualization overheads (due to the shared resources) are the main factors that affect remote display performance and the overall Quality of Experience (QoE), starting with the analysis made by Park and Kenyon in [6], which shows that network latency affects the timing characteristics of the provided service and, thus, the quality perceived by the user. This line of research was later on followed by Simoens et al. in [13], where the authors noted the correlation between the network data packetization overhead and the delay of user input events in remote display. To reduce this delay, the paper proposed an input buffering approach, leading to bandwidth savings up to 78%, while keeping the average user perceived responsiveness below 150ms. Triyason et al. in [35] also identified the impact of the remote display screen size on the user QoE in cloud-based environments.

A number of other studies [14,15,16] benchmarked how network latency impacts on the perceived performance in multiplayer Internet games, more specifically how delays in interaction affect the movement, location, and actions of a player. Three implementations of remote display protocols (Microsoft RDP, Teraddici PCoIP and Blast) were investigated in [7] to compare the screen image on the VDI client side, which is what the end user actually sees, with the screen on the remote side, given two different network speeds. The authors proposed a quality score and identified that the implementation of RDP and network speed in VDI environment both had an impact on the quality degradation. On the authentication performance, the study by Antal et. al [17] measured the impostor detection performance of a proposed machine learning model focusing on mouse dynamics using a dataset collected from a real organization, but the experiment was LAN-based and the network characteristics of the VDI were not taken into account. The study aimed to identify illegally used accounts by using time-series data of mouse movements of multiple users working in VDI. It showed that behavioural biometric technology based on mouse dynamics may be used as part of intrusion detection systems.

The area of keystroke authentication has been extensively investigated by prior research, but the main focus has been system-based. This is partially to do with the maturity of the concept, which was proposed and researched in the 2000-2010 decade, prior to the expansion of cloud/remote connectivity and access. A number of studies therefore evaluated existing algorithms, aimed to evaluate the entropy quality of the inputs and improve it through additional processing, comparatively analyzed existing algorithms, or highlighted limitations of the encompassing techniques. On the data entropy, M. Lizarraga in [18] collected keystroke dynamics from a set of the users, created template for each user and then tested the similarity across templates and samples to determine the efficiency of the verification algorithm when user is legitimate and when user is an impostor. The method of verification of the similarity was developed by authors and based on mean and standard deviation of each feature. The three key interaction features extracted (down-down, down-up, and up-down times) were a variant of the ones mentioned in Fig.1. Further, in [19], the authors created a benchmarking dataset from 16 different users and evaluated five different authentication algorithms. The raw input (consisting of a fixed password across three sessions) was used to build a biometric template (enrolment vector) which was then compared with two test vectors against the investigated algorithms. The template included only key timings information (also similar to the ones summarized in Fig.2) but, as the experiment was system-based, no network or environmental conditions were taken into account. Similarly, [20] observed the timing values from 10 users when entering the same password/username pair. The flight and hold time for each user were analysed and showed the difference of the user's biometric behaviour, some users typed text faster, some slower. Variations of this approach were attempted later on by [21, 22] by using timing characteristics of the keystroke as pressing time, dwell time and total time of password entering and calculating mean times for the respective interactions. If the value is above the threshold value, then access can be grated, otherwise access is denied. In both papers the users were required to enter only password for collecting biometric pattern.

In parallel with fixed-text methods, the studies summarised by [23] and [24] evaluated non-static biometric techniques based on free-text typing dynamics in order to determine their applicability for continuous authentication. The user typing profiles were obtained using of digraph-specific measures, and in [23] Impostor Pass Rate (IPR) and False Alarm Rates (FAR) indicated that the technique may indeed be applied as part of a multi-factor authentication system; also, the dynamic calculation allows it to work effectively even when a long time has passed since the reference user profiles were formed.

Studies by Gunetti et al. [23] and Araujo et al. [18] used distance, measured in deviation units between the pattern and the sample, to evaluate and discuss performance metrics. Gunetti normalised the means and standard deviations for distances between keystroke samples to [0;1] float values in order to understand how each of them can be used to separate keystroke samples, as well as their influence on FAR and FRR [23]. Araujo followed a similar approach in the [0;10] interval and visually demonstrated the separation between keystroke samples of two subjects [18]. Both studies showed that difference between average of distances measure of two users is at least 10%. Given their conclusion, if experiments indicate an impact of RDP latency to distance metric is greater or equal to 10%, then we may conclude that FAR, FRR would be also affected.

The entropy of keystroke data was further investigated in [25], more specifically the feasibility of using keystrokes for creating cryptography key for files stored remotely in the cloud as alternative to the algorithmic key generation. The key based on the biometry provides both authentication and integrity of data. The work showed the reliability of creating a 256-bit key using a biometric pattern – while the noise of the data led to a relative weak absolute accuracy, the results also showed very low FAR which made the method appropriate for usage as a secondary method for MFA.

The success of keystroke authentication depends on the stability of typing characteristics over time. In order to determine how timing variations affect its effectiveness, [26] presents the results of static manipulation with a dataset of collected keystroke dynamics. The time related features were extracted from the dataset, VKF is calculated based on the typing speed. Using Genetic Algorithm and Support Vector Machines (SVM) algorithms the feature subset selection was done from the reduced. Feature reduction rate and FAR were calculated using the MATLAB.

With regards to cloud-based usage of biometric authentication, in [27] the authors used 36 features to form a user profile. Keystroke activity was one of the input parameters for creating a user template together other host-based characteristic and network flow-based features, but only two features were used to describe keystroke activity of the user: the key press-down time and the time interval between two pressed keys.

To conclude, there have been significant research efforts to determine the feasibility, stability, and applicability of using keystroke dynamics for authentication. However, there have been only isolated, limited attempts to investigate how network impairments, delay and packet loss in particular, may affect the effectiveness of authentication using free-text keystroke biometrics. This paper aims to improve on the prior research by investigating the impact of network and host characteristics on the user biometric profile based on keystroke dynamics and to derive the threshold for delay variation beyond which the underlying behavioural biometric algorithms become ineffective.

## IV. METHODOLOGY

According to the NIST information security guidelines [28], the deployment of information system components with minimal functionality, such as thin clients, potentially reduces the likelihood of information exposure and successful attacks, hence are indeed a good option for maintain a security perimeter while also providing remote access. However, from a security architecture perspective, a thin client, given it resides outside the security perimeter of the organization, it is not required to have built-in security controls for authentication. This does create a vulnerability as compromised thin clients may lead to attackers gaining access mid-session to the remote desktop host, unless authentication is continuously validated. In this context, the necessary VDI scenario is one where the system performs continuous, transparent authentication and user access control on the Virtual Remote Desktop side, as depicted in Fig.1. This requires the keystroke pattern to be captured and verified on the remote session computer rather than on the thin client computer.

As mentioned in the literature survey section, a number of prior studies, including [26] identified that variations in the characteristics of typing do affect the accuracy of keystroke dynamics. However, the approach of existing research was to analyse the variations due to inconsistency in user behaviour rather than external parameters. We are therefore proposing to investigate whether variation in network latency can introduce noise into the keystroke timings to the extent that it would render it ineffective. The null hypothesis of this study is that network latency has no effect on the consistency of behaviour pattern calculated based on keystroke dynamics of user input. As part of the experimental approach, we used the input rendering engine for thin client, implemented several algorithms to calculate biometric pattern from real input keystroke dynamics in a VDI session, and then calculated the Euclidean distance between the user reference and current biometric pattern. In order to comprehensively investigate the impact of the external factors, we aim to take into account network protocol, network load, running processes on the remote host and observe their effects on the resulting keystroke dynamics biometric pattern and subsequently on the verification of the user profile. We acknowledge that other methods, such as machine learning algorithms or neural networks, may have different accuracy results versus Euclidian distance, but including actual identification or verification process is outside the scope of this work.

### A. Dataset

For comparative and reference purposes, the dataset of natural human-computer interaction chosen for the analysis was dataset from [29] which includes keystrokes dynamics of 39 users and three typing patterns of each user, resulting into a total of 122 typed keystrokes to be used as input data for creating user profile. This reference profile will be compared with current user profile that is captured under different network and host performance. The dataset contains keystroke-timing information from three different passwords, free-text questions (500 chars) and the transcript of Steve Jobs' Commencement Speech split into two parts. Each user performed the typing test in two separate laboratory sessions, with each session taking about one hour and containing approximately ten thousand keystrokes. Each keystroke sample is represented as $K_{u,j} = \{k_1(u,j), k_2(u,j),..,k_n(u,j)\}$ where $u$ is a user, $j$ is a number of samples; $n$ is an amount of the keystrokes in the sample $j$ of the user $u$. Each file in dataset contains multiple records with user ID and three features. Each keystroke $k_i(u,j)$, where $i \leq n$ included the following parameters:

1) user identifier – an index $u$, where $u = \overline{1,39}$, representing the user ID;
2) key state of the key – boolean k, where 0 means key pressed; 1 means key released;
3) key code - $c_i(u,j)$ represented in ASCII code;
4) timestamps - $td_i(u,j)$ and $tup_i(u,j)$ key press or release respectively by "a" feature, expressed in miliseconds.

The pressing and releasing of the keystroke can be represented as ($u$, "0", $c_i$, $td_i$) and ($u$, "1", $c_i$, $tup_i$) respectively, where $i$ is keystroke.

### B. Keystroke Replication

The dataset from [29] was used as input to use a reference input and to allow comparison of the results with the prior research. For consistency purposes, the dataset samples were replayed in order to emulate user typing on the thin client side but then transferred using RDP to the remote machine. We used the parameters from section above as raw input and wrote a small routine to generate input events by using the Windows API *keybd_event* function. We randomly selected an user and random sequence of 122 typed keystrokes that represents about 15 seconds of typing. The dataset samples of this randomly selected user were therefore replayed 40 times on host machine, transferred by RDP to the target remote machine, then the timing measures of the typing events was recorded into a new dataset using another Windows API routine. During the transmission of the text, a number of network- and system-based applications were run to emulate concurrent activity. The approach used may seem over-engineered in comparison to introducing noise directly into the input timing sequence, but we aimed to have an accurate end-to-end emulation of the network effects rather than a synthetic addition of them; also, there have been very few studies looking at how accurately keystrokes timing is reproduced following RDP transmissions through network environments with variable levels of congestion

### C. Feature Extraction

In order to verify the impact of external impairments on the effectiveness of keystroke authentication, we selected a recognised, successful method and applied it to both the raw initial dataset and the new dataset, based on the received keystroke events. We followed the studies [18, 30] and used digraph latency for feature extraction. Features were computed for each key-pair ($k_i$, $k_{i+1}$) using two main values, specifically the press time (P) and the release time (R) of each key in milliseconds. Four keystroke features were extracted from each key-pair for each user as shown in Fig.3:

1) *Press-Release Time:* Hold time for a key $k_i$ that define the time the key remains pressed and represented as a vector $PR_i(u,j) = \{pr_1(u,j), pr_2(u,j),..,pr_n(u,j)\}$, where $pr_i(u,j) = tup_i(u,j) - td_i(u,j)$ for $i \in n_{j,u}$
2) *Press-Press Time:* Time interval between pressing two successive keys ($k_i, k_{i+1}$). This feature is represented by $PP_i(u,j) = \{pp_1(u,j), pp_2(u,j),..,pp_n(u,j)\}$, where $pp_i(u,j) = td_{i+1}(u,j) - td_i(u,j)$ for $i \leq n$, where $n$ is an amount of the keystrokes in the sample $j$ of the user $u$.
3) *Release-Release Time:* The interval between releasing two successive keys ($k_i, k_{i+1}$), a composite keystroke latency feature defined as $RR_i(u,j) = \{rr_1(u,j), rr_2(u,j),..,rr_n(u,j)\}$, where $rr_i(u,j) = tup_{i+1}(u,j) - tup_i(u,j)$.

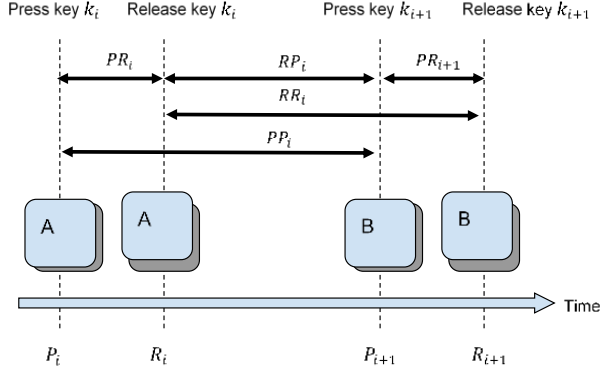

Fig.3. Timing features for the key pair

4) *Release-Press Time:* The interval between releasing and pressing two successive keys $(k_i, k_{i+1})$, and is represented as $RP_i(u,j) = \{rp_1(u,j), rp_2(u,j), \ldots, rp_{n-1}(u,j)\}$, where $rp_i(u,j) = td_{i+1}(u,j) - tup_i(u,j)$. This value could be positive or negative according to when keystroke $k_{i+1}$ was pressed, before or after $k_{i+1}$ was released. That is why we will use absolute value of this feature $|RP_i|$.

It is apparent only three of the four parameters are independent but, given there was no post-processing applied to the data, the full set of four was maintained for completeness.

Before feature extraction, the dataset was ordered by grouping records from dataset according to the keystroke value $c_i$ that first pressed and then released. The adjacency class was introduced as an input to set the level of adjacency between two keys on the keyboard followed by the study of Alsultan and Warwick [23]. This adjacency class allows to decrease the deviation of the time values during feature extraction. Each row was labelled by the class parameter that corresponds to one of the five adjacency classes and related to key-pair $(k_i, k_{i+1})$. Then, a row in the ordered dataset will have view ($c_i$, $td_i$, $tup_i$, *label*). This ordered dataset was used for the extraction of the timing features presented earlier and in Fig.4. For each feature (PR, PP, RR, RP) the mean value was calculated in accordance with the adjacency class to what the key-pair $(k_i, k_{i+1})$ was belonged.

Using the extracted features of the reference template $K(u,j)$, a user pattern is created for a particular typing session, consisting of mean time characteristics for 121 digraphs (corresponds to 122 characters). Each vector consists of 20 float values of the respective features.

$$\overline{K(u,j)} = \begin{pmatrix} PR_1 & PP_1 & RR_1 & RP_1 \\ PR_2 & PP_2 & RR_2 & RP_2 \\ .. & .. & .. & .. \\ PR_i & PP_i & PR_i & RP_i \end{pmatrix} \quad (1)$$

where $j$ is number of template samples (in our case j=1), $i$ is adjacency class ($i = 1..5$).

The current template $K'(u,j)$ is calculated the same way on the remote machine as reference template. To compare the similarity of two templates $K(u,j)$ and $K'(u,j)$, we calculate the Euclidean distance $\Delta KK'$ between the two vectors across the X=20 dimensions:

$$\Delta KK' = \sqrt{\sum_{x=1}^{X}(K(u,j) - K'(u,j))^2} \quad (2)$$

There are two approaches to calculate distance with several features for the authentication purpose. First way is to calculate distance for summative vector with all features that corresponds to the user profile, another way is to calculate distance value for each feature separately. The determination of the threshold is an important issue in the methodology. The analysis of the real collected keystroke data needs to be done to determine thresholds. The earlier works [18,30] showed that a user's feature with a higher variation demands a lower threshold, while a feature with a lower variation demands a higher threshold. So, the threshold for each feature in each account is obtained based on its standard deviation. Decision about an optimal threshold for each feature based on the a priori knowledge of authentication system administrator and usually the task of determination of the threshold is simple linear function maxima problem. In our work we do not consider decision-making part of the matching phase, the user's identification and authentication is out of scope. We only use distance value as indicator for changes that happened with keystroke pattern under certain conditions with variety of network and host characteristics.

V. EXPERIMENTS

To identify how traffic intensity and remote host characteristics influence the keystroke template vector, we implemented a testbed that emulates congestion and associated network impairments, consisting of two computers communicating via the Microsoft RDP protocol, which was chosen due to its popularity and availability. Two machines, acting as RDP client and RDP server, were connected via two routers over a serial link with a variable speed, adjustable via the bitrate of the link. The keystrokes were introduced on the RDP client from the dataset described above and the keystroke event timings were then collected on the RDP server. We ran the D-ITG traffic generator, described by [31] and available for download from [32], on the testbed to introduce variable traffic intensity that influences routers queues utilization [33]. D-ITG controls the amount of generated traffic by varying two parameters: C, which represents the intensity of added traffic in packets/sec and c, which defines the packet size in bytes. The resulting additional traffic and implicit level of congestion do affect the timings of the received keyboard events to the extent that they may be significantly far, from a Euclidian distance perspective, from the original events.

*A. Types of Experiments*

Three types of the experiment to introduce variable network conditions were conducted as part of the methodology: cross-traffic leading to network congestion and increased network latency, transmission of files between the remote desktop server and the host, affecting the RDP host processing, and streaming a video on the remote desktop, which impacts both the CPU availability on the RDP host and the network characteristics. In the first type of experiment, the transmission channel was loaded with constant bit rate (CBR) UDP concurrent traffic; the traffic consisted of two streams of the equal bitrate and the channel capacity was reduced to 1mbps. To establish the baseline behaviour, the network characteristics were initially measured in the absence of cross-traffic, then a first round of experiments was performed by varying the C parameter in D-ITG; three separate experiments were included - first gradually between 10 and 100 packets/second, then fixed at 110 packets/second and 115 packets/second. A separate round of experiments included a combination of traffic and host-based activities, experiment 5 corresponds to the conditions of the experiment 3 but with some activity on the remote display such as random fast moving of window with open application, in our case, moving of open window with File Explorer, then a parallel file download and video streaming. The file download and video streaming were run over 100Mbps network. For the file download, a large file was downloaded from the RDP server by the RDP client and video streaming used a fullHD Youtube video being played on the RDP server and viewed on the RDP client. A summary of the conditions for each experiment is presented below in Table I.

TABLE I
EXPERIMENTS SETUP

| Experiment | 1 | 2 | 3 | 4 | 5 | 6 | 7 |
|---|---|---|---|---|---|---|---|
| C parameter D-ITG [pkts/s] | n/a | 10-100 | 110 | 115 | 110 | n/a | n/a |
| Bottleneck capacity [Mbps] | 1 | 1 | 1 | 1 | 1 | 100 | 100 |
| Additional activities | - | - | - | - | Window movement | File download | Video streaming |

To ensure the statistical significance of the results, each of the seven experiments was repeated 40 times and the results were averaged. While the range of replicated conditions and environments does not provide an exhaustive list of impairments, it provides a number of typical cases that users are likely to encounter as part of their RDP interaction.

*B. Results*

In order to determine the impact of cross-traffic and host-based activities, we calculated the Euclidean distance between the initial vector and the vector collected at the RDP receiver, as affected by the respective interactions. The purpose of this study is not to provide a binary evaluation of specific algorithms or their behaviour when faced with impairments, but to determine the level of impact of these impairments on the keystroke dynamics. Given the majority of keystroke authentication methods are ultimately reverting to an evaluation of how close the measured sequence is to the stored profile, we feel that the Euclidian distance, while it not a complete match for a specific method, defines the level of this impact.

Table II below summarises the results of the experiments performed.

As shown by the first three experiments, the impact of low or medium congestion is minimal on the resulting keystroke dynamics parameters, as the Euclidean distance of experiments 2 and 3 is comparable to the baseline obtained in experiment 1. In the case of experiment 3, the measured average bitrate was 901 kb/s; this is close to the bottleneck bandwidth for the testbed, but the results indicate that it had minimal impact on the Euclidian distance between the original vector and the one collected at the RDP server. Although the channel utilisation was rather high, there was no packet loss recorded and the average delay (10.5ms) and jitter (3.2ms) were also low. In order to see the behaviour of the channel under heavy congestion, the data flows were further increased to an average bitrate of 916kb/s, leading to a bottleneck link utilisation of 91.7% and subsequently increasing the average delay (370ms); given the additional traffic was CBR, the jitter varied only slightly, increasing to 3.7ms. The overall dynamics of the distance value when varying the network latency parameters are summarised in Fig.4.

TABLE II
EXPERIMENTAL RESULTS

| Experiment | 1 | 2 | 3 | 4 | 5 | 6 | 7 |
|---|---|---|---|---|---|---|---|
| Av. Euclidean distance | 1.068 | 1.07 | 1.074 | 3.2 | 1.77 | 2.12 | 3 |
| Av. packet loss [%] | 0 | 0 | 0 | 3.6 | 0.14 | 0 | 0 |
| Average delay [ms] | <1 | <10 | 10.5 | 370 | 64 | n/a | n/a |

The last three experiments involved a combination of RDP-based interaction, RDP-based traffic, and bandwidth limitations. Experiment 5 was similar to the original conditions, running through a limited 1Mbps bandwidth, but included additional activity on the remote display, more specifically random fast window movements for an open application. This affected slightly the average delay, increasing it to 64 ms, but had a slightly more significant impact on jitter (7ms) and packet loss, which increased to 0.14%. While none of the individual parameters increased significantly, the resulting Euclidian distance varied by a significant factor of 1.77.

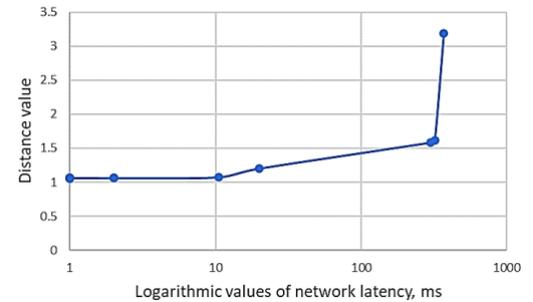

Fig.4. Performing the similarity of keystroke dynamics pattern under different network latency.

Experiments 6 and 7 also investigated the impact of additional RDP activities onto the keystroke timings, but using a non-restricted 100Mbps link. For experiment 6, a text file was transmitted from the remote machine to the client machine via RDP, while experiment 7 generated additional traffic by streaming a video from the remote machine to the thin client. Both activities could potentially be equally disruptive; TCP is the transport protocol for bulk data transfer and is meant to be a conservative and network-friendly, but also has the ability to expand to the available bandwidth in order to maximise its efficiency. The video chosen for experiment 7 was an HD 1080p video, with bitrates varying between 4000-8000kbps (https://www.youtube.com/watch?v=AWKzr6n0ea0); while this is indeed the bitrate between the YouTube server and the RDP server, optimised for the bandwidth between the two endpoints, not the same can be said about the link between the RDP server and the client, which uses non-optimised video communication between them. As a result, it is likely that the down-streaming from the server to the client is higher in terms of bitrate.

To determine the significance of results for experiments, we performed series of tests where we calculated the Euclidean distance between the initial vectors of two random users $K(u)$, $K'(u)$ collected at the thin client computer after keystroke replication process. The average value for Euclidean distance was equal to 8.5 for two absolutely different users. If we compare the result with the resulting distance after experiment 1 without additional traffic, we may conclude that VDI testbed adds 12% of distortion into the distance measure of synthetic keystrokes, high network latency (experiments 3-5) adds 15%-20% of distortion, and RDP overheads (experiments 6 and 7) add 28%-38% of distortion to Euclidean distance.

### C. Discussion

The results showed that increasing the channel load with UDP traffic to a certain level does not affect the biometric pattern. This could be explained by the fact that UDP traffic is non-bursty, smooth, and the standard deviation and jitter values are relatively small. But when the channel utilization goes beyond 0.9 (as it is the case for experiments 3-5) the throttling algorithm is starting to work causing a packet loss. This affects the keystroke pattern and the similarity coefficient changes from 1 to 1.7 even with small packet loss values of 0.15-0.5%.

The keystroke dynamics pattern is also affected by the value of standard deviation and jitter, as demonstrated where the transmission of RDP keyboard events is concurrent with events from the VDI. In this case, the average delay was within normal limits but the deviation in delay values was almost two times greater (100 msec). From this, we can conclude that such an effect will be caused by applications running on TCP, since TCP traffic is characterized as bursty, greedy, and is more resistant to delays, but has large jitter values due to its nature.

When the network load approaches bandwidth the network latency increases to an unacceptable 370 msec (ITU-T recommendation for UDP traffic network latency is less than 150 ms) and packet loss occurs. This leads to a modification of the timing information required for user current biometric pattern, as a result, the similarity coefficient has a very large distance value which means that the user is most likely not to be identified.

Additional actions such as copying and watching videos on the remote display also significantly affect the keystroke pattern which means that biometrics dynamics are transmitted with the same priority as the rest of the traffic and are served in a queue. From this we can conclude that continuous authentication will be not effective in this situation.

Not prioritizing of RDP messages that carry biometric information can be explained by the fact related to the functioning/processing of keyboard events. Keyboard messages often have a low priority over other operations such as file management [34]. This is apparent when typing in a word processor. If there is any disk activity when typing, the text being written is held in a buffer until such time that the processor is free to deal with it (causing the display to pause and then the characters appearing). In order to achieve the high accuracy of timing between keystrokes, the keyboard driver and timing device need a high priority in order not to be affected by other activities.

## VI. CONCLUSION

This paper investigated the impact of network traffic and transport infrastructure constraints on the behaviour of the keystroke pattern in the remote desktop applications and VDI. The variation in the keystroke pattern was estimated by measuring the Euclidian distance, which is at the core of matching algorithms that are based on the distance between reference user profile and the current user profile.

A raw dataset was converted to a more descriptive set of four timing features (PR, PP, RR, and RP) that was used as input for the experiments. This resulting profile was then compared with the timing patterns obtained when exposing the RDP communication channel to increasing network latency or/and additional traffic, such as transmission of files from the remote desktop to the host or streaming a video from the remote desktop. In order to avoid bias due to the adaptiveness of the underlying biometric algorithm, the resulting profiles were compared using Euclidean distance that is a metric in instance-based learning algorithms such as k-nearest neighbours (KNN) algorithm.

Following from the results of the experiments, we found that the user timing pattern is not affected by network latency, but standard deviation, jitter, and packet loss have a significant combined impact onto it. This suggests that, in spite of the apparent network-unfriendly nature of UDP, working with applications that require TCP transmission on the remote desktop is likely to have a greater effect on the pattern than UDP based ones. It can also be concluded that RDP does not prioritise its encompassing data flows, which means that any competing traffic is likely to render keystroke dynamics unusable for continuous authentication for remote access. For one-time authentication at the beginning of the session it is possible to use keystroke pattern, as long as there are no other running applications or open windows that can affect the timing.

In a continuous authentication system using behavioural biometrics, the key role of the acquisition module is to ensure one of the four requirements to biometric data - permanence and immunity to circumvention for the keystrokes. As this study demonstrates, the combination of the RDP protocol and a typical network environment may have an effect on this requirement, which subsequently may impact the FAR, FRR parameters of the biometric systems based on keystroke dynamics.

Although it is a small part of it, keystroke is one the two (together with mouse movements) inputs for RDP and therefore the traffic that carries it should be prioritised to ensure its timely arrival. At the time of writing the article, spring 2020, the importance of remote working and access has become more evident than ever before and, subsequently, biometric identification and authentication using keystroke dynamics could represent a great solution to improve the security and control of remote access.

For future work, we plan to focus more on studying the influence of network characteristics, such as standard deviation and jitter on the keystroke dynamics pattern, as well as TCP traffic. We also plan to experiment with other remote display protocols and remote desktop applications as AWS PCoiP, and Citrix ICA, Microsoft Teams.

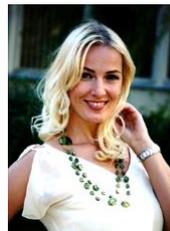

**Ievgeniia Kuzminykh** received the Ph.D. degree in telecommunications in 2013 from the Kharkov National University of Radio Electronics where she is currently a Visiting Associate Professor with the department of infocommunication.

From 2017 to 2020, she was a Senior Lecturer with the computer science department in Blekinge Institute of Technology, Sweden. She was holding joint appointment as postdoc researcher with the Technical University of Denmark in 2019-2020. From July 2020 she is Lecturer in Cybersecurity Education with King's College London. She has coauthored over 40 publications. Her research interests include cybersecurity, IoT, security aspects of cloud and networks.

Dr. Kuzminykh has served for various conferences and journals as a reviewer (IEEE Trans. on Cloud Computing, Hindawi Journal of Security and Communication Networks, MDPI Journal of Applied Sciences, Acta Innovation Journal, IEEE conference Problems of Infocommunications Science and Technology).

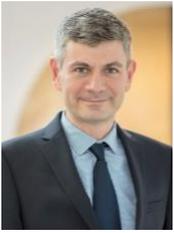 **Bogdan Ghita** received his PhD in 2005 from University of Plymouth, UK. He is Associate Professor at University of Plymouth and leads the networking area within the Centre for Security, Communications, and Network research. His research interests include computer networking and security, focusing on the areas of network security, performance modelling and optimization, and wireless networking, published over 150 papers, graduated 20 PhD students, and having been principal investigator in a number of industry-led, national, and EU research projects in these areas.

He was a TPC member for over 100 international conference events as well as a reviewer for IEEE communications letters, computer communications, and future generation computer systems journals.

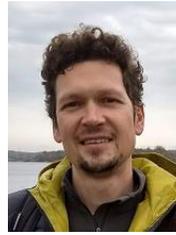 **Alexandr Silonosov** is currently pursuing his M.S. degree from the department of computer science from Blekinge Institute of Technology, Sweden, where he is jointly working as a Teaching Assistant. Last ten years he is working on authentication, identity management and user behavior auditing solutions. His research interests are designing and building desktop and network security systems, multi-factor authentication frameworks, ML in information security.

He is holder of certificates: Microsoft Certified Professional. Designing and Implementing Desktop Applications with Microsoft, I2C CISSP.